\documentclass{ws-procs961x669}
\usepackage[super,compress]{cite}
\usepackage{graphicx}
\usepackage{amsmath,amssymb}
\usepackage{mathtools}

\newcommand{\tr}[2][]{\textnormal{tr}#1{{\left\{#2\right\}}}}

\newcommand{\Eq}[2][Eq.~]{#1(\ref{eq:#2})}
\newcommand{\half}{\frac{1}{2}}
\newcommand{\thalf}{\tfrac{1}{2}}
\newcommand{\D}{\mathrm{d}}\newcommand{\I}{\mathrm{i}}
\newcommand{\Exp}[1]{\mathrm{e}^{\mbox{\footnotesize$#1$}}}
\newcommand{\ns}{n_{\mathrm{s}}}
\newcommand{\nr}{n_{\mathrm{r}}}
\newcommand{\Ns}{N_{\mathrm{s}}}
\newcommand{\zetas}{\zeta_{\mathrm{s}}}

\DeclareMathAlphabet{\vecfont}{OT1}{cmr}{bx}{it}
\renewcommand{\vec}[1]{\vecfont{#1}}
\newcommand{\grad}{\boldsymbol{\nabla}} 

\makeatletter
\renewcommand{\ps@plain}{%
  \renewcommand{\@oddhead}{\hfil{\footnotesize%
    A contribution to the Julian Schwinger Centennial Conference, %
    7--12 February 2018, Singapore}\hfil}%
  \renewcommand{\@evenhead}{\@oddhead}%
  \renewcommand{\@oddfoot}{\hfil\thepage}%
  \renewcommand{\@evenfoot}{\thepage\hfil}%
}
\makeatother
\crop[off]

\begin{document}

\title{The Statistical Atom%
  \footnote{This review was written in 1985 as an invited article for
    Physikalische Bl\"atter, but the editor did not like what we submitted
    and did not put the paper into print.
    A few years later, it was to be the very first paper in a new 
    physical chemistry journal, which never came into existence.
    After decades in limbo, the review is now a fitting contribution to these
    proceedings.}}

\author{Julian Schwinger and Berthold-Georg Englert%
  \footnote{Now at Centre for Quantum Technologies and at Department of
    Physics, National University of Singapore; and at MajuLab, Singapore.}} 

\address{Department of Physics, University of California, %
         Los Angeles, CA 90024}

\begin{abstract}
This review is the updated and enlarged version of a talk delivered by
J.~S. on the occasion of the 1982 meeting of Nobel laureates at Lindau, and of
talks given by \mbox{B.-G.~E.} at several West German universities and Max
Planck institutes in 1984. 
\end{abstract}

\section{Introduction}
The title of this review indicates the two main themes of the subject.
The generality of ``\emph{The} Statistical Atom'' emphasizes the ambition of
dealing, not with a specific chemical element, but with the Periodic Table as
a whole. 
And the word \emph{statistical} points to the method applied in these
investigations.
Statistics has two different meanings here.
First, many-electron systems obey Fermi--Dirac statistics, of course.
Second, and more to the point, some properties of atoms can be studied by
looking first at situations involving large numbers of electrons. 

Let us supply some evidence for the practicality of such a statistical
approach.
The most primitive theoretical model neglects the inter-electronic
interaction, thus treating the electrons as independently bound by the
nucleus. 
But even if fermions do not interact they are aware of each other through the
Pauli principle.
Therefore, such noninteracting electrons (NIE) will fill the successive Bohr
shells of the Coulomb potential with two electrons in each occupied orbital
state.
Since the degeneracy of the shell with principal quantum number $n$ is
$2n^2$-fold, the total number, $N$, of electrons in $\ns$ filled shells is 
\begin{equation}\label{eq:1}
  N= \sum^{\ns}_{n=1}2n^2
  =\frac{2}{3} \biggl( \ns + \frac{1}{2} \biggr)^3
  - \frac{1}{6} \biggl( \ns + \frac{1}{2} \biggr)\,.
\end{equation}
The total binding energy for a nucleus of charge $Z$ is even simpler,
\begin{equation}\label{eq:2}
-E= \sum^{\ns}_{n=1}2n^2 \frac{Z^2}{2n^2}=Z^2 \ns\,,
\end{equation}
which uses the single particle binding energy $Z^2/(2n^2 )$.
[Here and in the sequel we adopt atomic units, which measure energies in
multiples of twice the Rydberg constant, $me^4/\hbar^2 = 27.21\,$eV, and
distances in multiples of the Bohr radius, $a_0= \hbar^2 /(me^2)=0.5292\,$\AA.]
If we now invert \Eq{1} for large $N$, we get the asymptotic energy formula
\begin{equation}\label{eq:3}
  \frac{-E}{Z^2}=\ns \cong \biggl(\frac{3}{2}N\biggr)^{1/3}
  - \frac{1}{2} + \frac{1}{12} \biggl(\frac{3}{2}N\biggr)^{-1/3} + \cdots\,.
\end{equation}

This is certainly a good approximation if the number of electrons is large
enough.
But suppose we apply it for a small number, say $N = 2$, where the exact value
of $\ns$ is one?
Well, this asymptotic formula produces
\begin{equation}
\ns \cong 1.0000297 \quad \textrm {for} \quad N=2
\end{equation}
which is in error by only 0.003\%!
For $N = 10$, the deviation from $\ns= 2$ is substantially smaller:
just 0.0001\%.
We take this practically perfect agreement as strong encouragement for trying
a similar large-$N$ approximation for realistic atoms, where the electrons do
interact.
But before going into those details, we point out that the primitive NIE-atoms
can supply realistic qualitative answers.
For example, the total binding energy of neutral systems ($N = Z$) can be
written as 
\begin{equation}\label{eq:5}
\frac{-E}{\frac{1}{2}Z^2}=2.289 Z^{1/3} - 1 + 0.1456 Z^{-1/3} + \cdots\,,
\end{equation}
a structure that we shall meet again for real atoms, but with somewhat
different numerical factors.
Then consider atomic size.
The dimension of a Bohr shell is specified by the square of the principal
quantum number divided by the charge of the nucleus.
For neutral atoms this says that
\begin{equation}\label{eq:6}
\textrm{atomic size} \sim \frac{\ns^2}{Z} \sim Z^{-1/3}\,,
\end{equation}
which again is qualitatively correct for the main body of electrons.

\begin{figure}[t]
\centering
\includegraphics{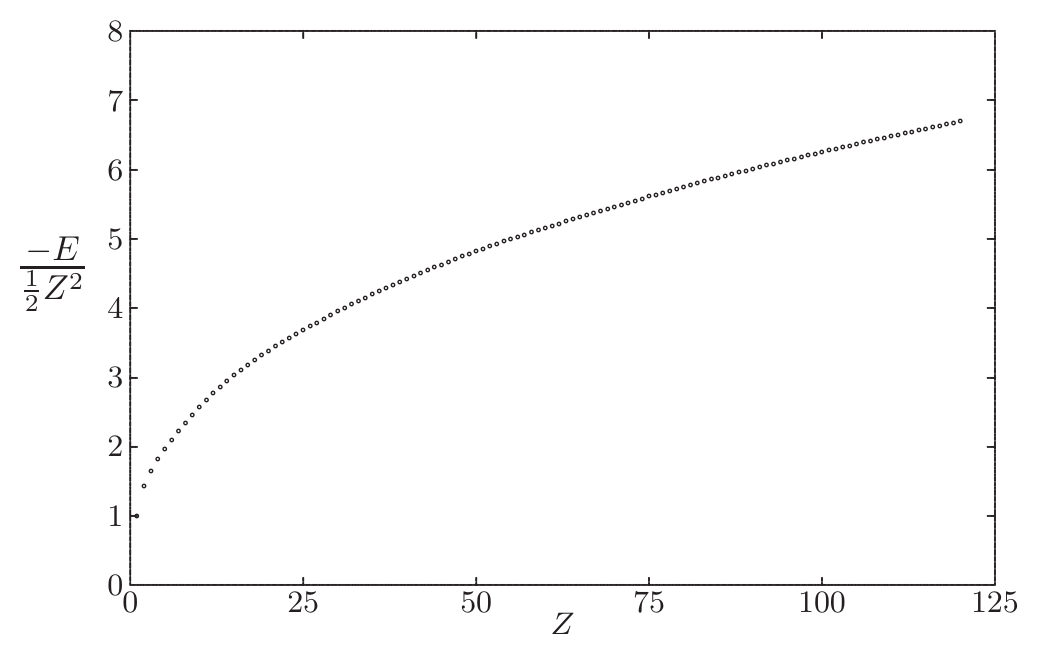}
\caption{HF binding energies for $Z=1, \ldots, 120$.}
\label{fig01}
\end{figure}

Another kind of support for the statistical, i.e., large-$N$ treatment of
atoms comes from a look at properties of real atoms.
A good example is the total binding energy, as presented in Fig.~\ref{fig01}.
Each individual circle in this figure represents the result of a Hartree--Fock
(HF) calculation,\cite{Desclaux} which energies agree reasonably well with
experimental values where they are available.
Yet there is no understanding supplied by these individual calculations, at
integer values of $Z$, for the fact that these binding energies behave so
remarkably regularly as a function of the atomic number $Z$.
Why then should one expect the large-$N$ approximation to be useful?
Simply because such a regular dependence on $Z$ (or $N$) cries out for a
formula like \Eq{5}, and the statistical approach is likely to produce it.

\section{General Approach}
The basic simplification provided by the large number of electrons in the
statistical atom is the possibility of introducing an average potential in
which the electrons can be considered to be moving independently.
That effective potential, $V$, describes both the interaction with the nucleus
and the electron--electron interactions. 
We thus start with the one-particle Hamiltonian (atomic units again)
\begin{equation}\label{eq:7}
H = \frac{1}{2} \vec{p}^2 + V\,,
\end{equation}
and use it to write down the total one-particle energy and the total number of
electrons when all states with energy less than $-\zeta$ are occupied, 
\begin{eqnarray}
E_{\mathrm{1p}}&=&\tr{H\,\eta(-H-\zeta)}\,,\nonumber\\
N&=&\tr{\eta(-H-\zeta)}\,.
\end{eqnarray}
The combination $H+\zeta$, that appears in the argument of Heaviside's step
function $\eta$, invites rewriting $E_{\mathrm{1p}}$ as
\begin{equation}\label{eq:9}
  E_{\mathrm{1p}} = \mathrm{tr} \{ (H+\zeta) \eta(-H-\zeta)\}
  - \zeta N \equiv E_1 - \zeta N\,.
\end{equation}
This sum of single particle energies counts every electron pair interaction
twice.
In order to obtain the real energy, we therefore have to subtract the
electron--electron interaction energy once.
If we disregard the exchange interaction for a start, this is just the
electrostatic energy of the electronic cloud.
It is most advantageously evaluated in terms of the integrated square of the
electric field.
Thus, the total energy is given by
\begin{equation}\label{eq:10}
  E= E_1 - \int(\D\vec{r})\, \frac{1}{8 \pi}
  \Biggl(\grad \biggl( V + \frac{Z}{r} \biggr) \Bigg)^2 - \zeta N \,,
\end{equation}
where $-Z/r$ is the potential of the nucleus, which has to be
subtracted from $V$ because only the field produced by the electrons is asked
for. 

The advantage of the particular combination of energies in \Eq{10} is its
stationary property under variations of $V$ and $\zeta$.
First notice that the response of $E_1$ to infinitesimal changes of $V$
exhibits the electron density $n$,
\begin{equation}\label{eq:11}
\delta_V E_1 = \int(\D\vec{r})\, n\, \delta V \,.
\end{equation}
Then we indeed find a vanishing first order change of $E$, i.e., $\delta_VE=0$,
in consequence of Poisson's equation 
\begin{equation}\label{eq:12}
n = - \frac{1}{4\pi} \grad^2 \biggl(V + \frac{Z}{r} \biggr)\,.
\end{equation}
Likewise, the variation of $\zeta$ produces no first order change, because
\begin{equation}\label{eq:13}
N = \frac{\partial E_1}{\partial \zeta} = \int (\D\vec{r})\, n \,,
\end{equation}
where the last equality is a simple consequence of the fact that $V$ and
$\zeta$ appear in $E_1$ only as the sum $V+\zeta$.
Equation~(\ref{eq:12}) is a differential equation for the potential $V$, while
\Eq{13} is an algebraic equation that determines $\zeta$.
We thus have just enough information to find both $V$ and $\zeta$ for given
$Z$ and $N$. 

Obviously, the essential difficulty in this general approach is the evaluation
of the trace in \Eq{9} for an arbitrary potential $V$, and then the subsequent
calculation of the density $n$, needed in Poisson's equation \Eq[]{12}.
Hartree's way of solving this problem is to write down the one-particle
Schr\"{o}dinger equation for $H$ of \Eq[]{7}; look for eigenvalues and
eigenfunctions; then square the wave functions to finally produce the density,
which in turn leads to a new potential to be used for the next iteration.
This method imitates the exact treatment of hydrogen.
But the idea of the average potential is best justified at the other end of
the Periodic Table, where there are many electrons.
So a different way of evaluating the trace of \Eq{9} is called for.
It was invented more than half a century ago by Thomas and, independently,
by Fermi.\cite{Thomas-Fermi}

\section{The Thomas--Fermi Model}
The Thomas--Fermi (TF) approximation is based on the following idea.
Although it is true that the potential $V$ changes substantially from deep
within the atom to far outside, it should not vary significantly over the
range important for a single electron, provided there are many electrons.
If that is so, it will be a reasonable approximation for $E_1$ to sum the
classical single-particle energies $\frac{1}{2}\vec{p}^2 + V (\vec{r})$ over
those cells in phase space that are occupied.
The counting is left to quantum mechanics; two electrons per phase space
volume of $(2\pi \hbar)^3$ [$=(2\pi)^3$ in atomic units].
Thus, we write
\begin{equation}\label{eq:14}
  E_1 = 2 \int \frac{ (\D\vec{r})(\D\vec{p})} {(2\pi)^3}
  \biggl( \frac{1}{2}\vec{p}^2 + V + \zeta \biggr)\,
  \eta \biggl( - \frac{1}{2} \vec{p}^2 - V - \zeta \biggr) \,.
\end{equation}
The step function $\eta$ cuts off the momentum integral at the
($\vec{r}$-dependent) maximal momentum,
\begin{equation}\label{eq:15}
P = \sqrt{-2(V+\zeta)}\,,
\end{equation}
so that
\begin{equation}\label{eq:16}
  E_1 = \int (\D\vec{r}) \biggl( - \frac{1}{15\pi^2} \biggr) P^5
  = \int (\D\vec{r}) \biggl(-\frac{1}{15\pi^2} \biggr) [ -2(V+\zeta)]^{5/2}\,.
\end{equation}
The density is found by differentiating the integrand with respect to $V$, in
accordance with \Eq[]{11}, thus producing, after insertion into Poisson's
equation \Eq[]{12}, 
\begin{equation}\label{eq:17}
  n = \frac{1}{3\pi^2} [ -2(V+\zeta) ]^{3/2}
    = - \frac{1}{4\pi} \grad^2 \biggl( V + \frac{Z}{r} \biggr) \,.
\end{equation}
This is the TF equation for $V$.
It has some simple but fundamental implications.

Far inside the atom, the potential is that of the nucleus,
$-Z/r$, large and negative.
Moving outwards the potential becomes less and less negative, finally equaling
$-\zeta$, after which the argument of the square root in \Eq[]{17} changes its
sign.
So there, at a certain distance $r_0$, the TF atom has, in general, a sharp
edge, beyond which the density is zero.
The picture is too simple to describe the exponential decrease of the density
in the outer regions of the atom.
At the edge, the potential is just the Coulomb potential of the net charge
$Z-N$, implying 
\begin{equation}\label{eq:18}
\zeta = \frac{Z-N}{r_0} = - V(r_0)\,,
\end{equation}
thus
\begin{equation}\label{eq:19}
\zeta = 0\quad \mathrm{for} \quad Z=N\,.
\end{equation}
The differential equation \Eq[]{17} requires an additional condition at the
otherwise undetermined distance $r_0$.
It is supplied by the continuity of the electric field, being the field of net
charge $Z-N$, or 
\begin{equation}\label{eq:20}
  \frac{\D V}{\D r} (r_0) = \frac{Z-N}{r_0^2}\,;
  \quad = 0 \quad \mathrm{for} \quad Z=N\,.
\end{equation}
The boundary conditions \Eq[]{18} and \Eq[]{20} are supplemented by the obvious
one at $r=0$, 
\begin{equation}\label{eq:21}
V \rightarrow - \frac{Z}{r} \quad \mathrm{for} \quad r\rightarrow 0\,,
\end{equation}
and together they specify a unique solution of \Eq[]{17} for every $N\leq Z$.
Equation \Eq[]{19} tells us that neutral TF systems are filled to the brim
with electrons.
There are no negative ions in this approximation.

Let us take a closer look at the TF potential for neutral systems.
It is useful to measure $V$ as a multiple of the potential of the nucleus by
introducing the TF function $F(x)$, 
\begin{equation}\label{eq:22}
V=-\frac{Z}{r}F(x)\,,
\end{equation}
the argument of which is related to the physical distance $r$ by
\begin{equation}\label{eq:23}
  x=\frac{Z^{1/3}r}{a}\quad\textrm{with}\quad
  a=\frac{1}{2} \biggl( \frac{3\pi}{4}\biggr)^{2/3}\,.
\end{equation}
The constant $a$ is chosen such that the differential equation for $F$,
\begin{equation}\label{eq:24}
\frac{\D^2}{\D x^2} F(x) = \frac{[F(x)]^{3/2}}{x^{1/2}} \,,
\end{equation}
also called the TF equation, is free of numerical factors. For neutral atoms,
$\zeta = 0$, the boundary conditions \Eq[]{18} and \Eq[]{20} can be satisfied
only at infinity, 
\begin{equation}\label{eq:25}
F(\infty) =0\,,\quad F'(\infty) = 0\,,
\end{equation}
and \Eq[]{21} translates into
\begin{equation}\label{eq:26}
F(0) = 1\,.
\end{equation}
Please notice that both the differential equation \Eq[]{24} and the boundary
conditions (\ref{eq:25}, \ref{eq:26}) do not refer to $Z$.
The TF function is a universal function for all $Z$ .
The potential $V$ itself does, of course, depend on $Z$; first through the
factor $Z/r$, but then also because of the $Z$ dependence of the
TF variable $x$ of \Eq{23}.
The factor $Z^{1/3}$ there implies the same shrinking of larger atoms that we
have already observed for NIE, see \Eq{6}. 

\begin{figure}[t]
\centering
\includegraphics{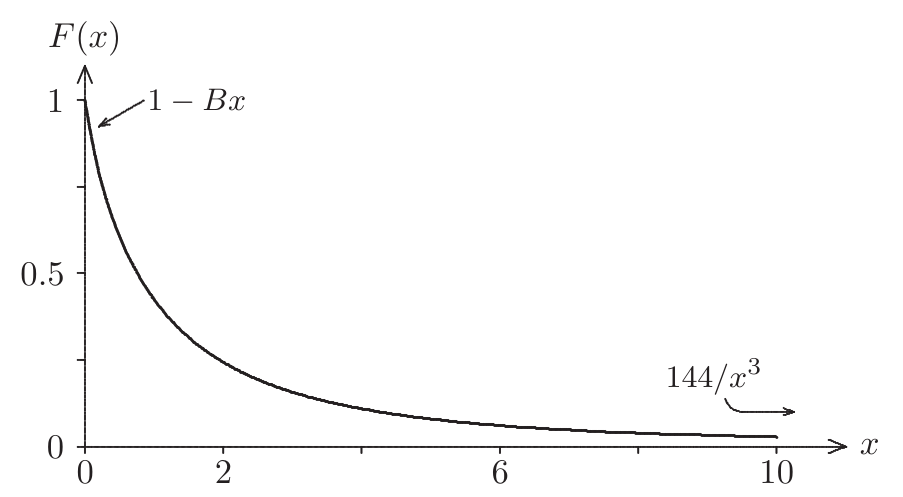}
\caption{The Thomas--Fermi function $F(x)$.}
\label{fig02}
\end{figure}

Figure \ref{fig02} shows a plot of $F(x)$, which is well known numerically.
For our purposes the initial slope $B$,
\begin{equation}\label{eq:27}
F(x) \cong 1 - Bx \quad \mathrm{for} \quad x \ll 1
\end{equation}
is important; its numerical value is (approximately)
\begin{equation}
B = 1.588\,.
\end{equation}
The physical significance of $B$ is apparent when \Eq[]{27} is inserted into
\Eq[]{22}, producing 
\begin{equation}\label{eq:29}
  V(r) \cong - \frac{Z}{r} + \frac{B}{a} Z^{4/3}
  \quad \mathrm{for} \quad r\cong 0\,,
\end{equation}
inasmuch as the additive constant is the interaction energy of an electron,
near the nucleus, with the main body of electrons.
We can use it to immediately write down the change in energy caused by an
infinitesimal change of the nuclear charge $Z$ to $Z+\delta Z$.
It is the analogous electrostatic energy of that additional nuclear
charge,\cite{fn3} where a minus sign is needed to connect with the known
energy, which is that of an electron,
\begin{equation}
\delta E = - \frac{B}{a} Z^{4/3} \delta Z \,.
\end{equation}
The consequence
\begin{equation}\label{eq:31}
E = - \frac{3}{7} \frac{B}{a} Z^{7/3}
\end{equation}
is the TF formula for the total binding energy of atoms.
Comparison of the numerical factor in
\begin{equation}\label{eq:32}
\frac{-E}{\frac{1}{2} Z^2} = 1.537 Z^{1/3}
\end{equation}
with the one of the leading term for NIE [\Eq{5}] shows that the
electron-electron interaction reduces the atomic binding energy by roughly one
third. 

\begin{figure}[t]
\centering
\includegraphics{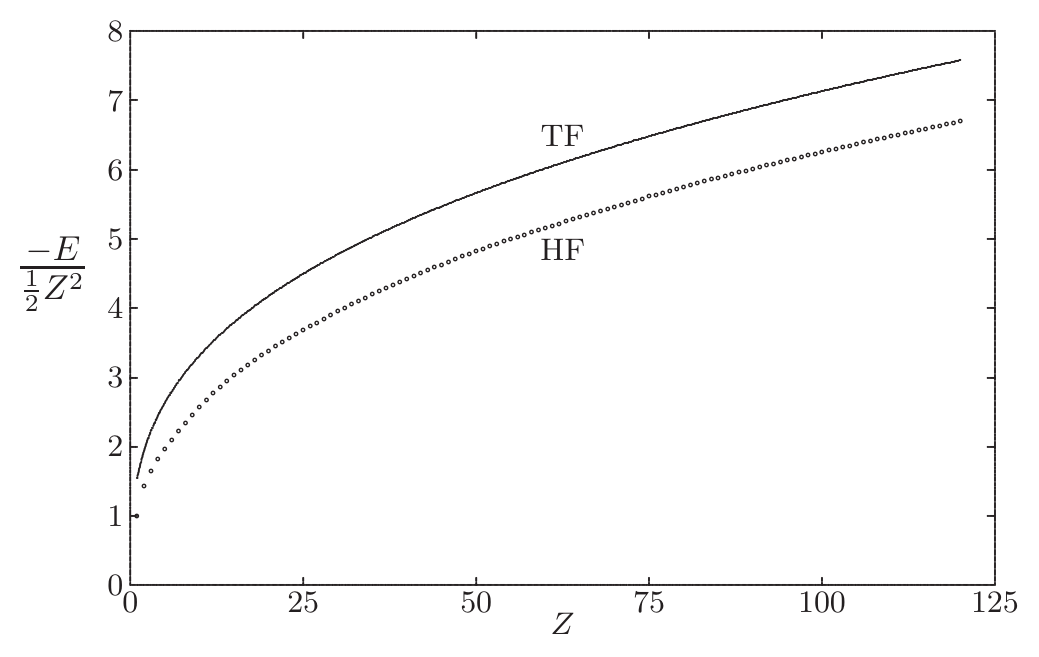}
\caption{Comparison of HF binding energies and the TF prediction, \Eq{32}.}
\label{fig03}
\end{figure}

A look at Fig.~\ref{fig03} shows that \Eq{32} does reproduce the general trend
of the atomic binding energies.
Although the need for refinements is clear, it is remarkable how well TF works
despite the crudeness of the approximation that it represents.
In Fig.~\ref{fig03} the continuous statistical curve is closer to the
integer-$Z$ HF circles at small $Z$ values than at large ones.
This is deceptive, however, since it is the fractional difference that
counts.
This relative deviation \emph{decreases} with increasing $Z$.
For $Z = 10$, $20$, $30$, $60$, $90$, and $120$ its amount is $29$, $24$,
$21$, $17$, $15$, and $13$ percent, respectively. 

\begin{figure}[t]
\centering
\includegraphics{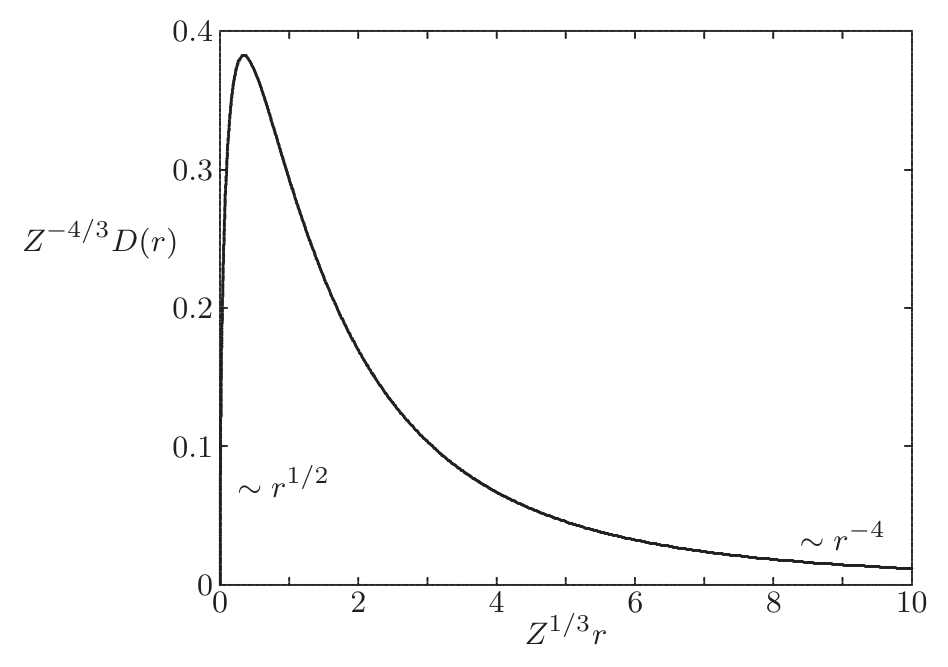}
\caption{Radial density $D = 4\pi r^2n$, as predicted by TF.}
\label{fig04}
\end{figure}

There are also obvious deficiencies of the simple TF model, graphically
demonstrated by a plot of the TF density in Fig.~\ref{fig04}.
At small distances the radial density grows like the square root of $r$, not
proportional to the square of $r$ as required by a finite density at $r=0$,
whereas the drop off at large distances goes like $r^{-4}$, which is so slow
that one never really gets outside the TF atom.
We have noted earlier that neutral TF atoms have their edge, $r_0$, at
infinity. 

\section{Validity of TF}
Before we can improve TF we must find out where it fails.
Recall that the derivation started with the assumption that $V$ is slowly
varying.
What does this mean?
The quantum standard of length associated with an individual electron is its
de Broglie wavelength, $\lambda$.
The potential does not change significantly over this range, if
\begin{equation}\label{eq:33}
  \boldsymbol{|} \lambda \grad V \boldsymbol{|} \ll
  \boldsymbol{|}V\boldsymbol{|}\,.
\end{equation}
Substantial changes in $V$ occur on a scale set by the distance $r$, so that
criterion \Eq[]{33} requires that 
\begin{equation}
\lambda \ll r\,.
\end{equation}
On the other hand, $\lambda$ is the inverse momentum (we ignore factors of $2$
or $\pi$ for this kind of reasoning), which in turn is given by the square root
of the potential, see \Eq{15}.
In short, we have, as criterion for the validity of TF, the relation
\begin{equation}\label{eq:35}
r \sqrt{\boldsymbol{|}V\boldsymbol{|}} \gg 1\,.
\end{equation}
Upon introducing TF variables, this reads
\begin{equation}\label{eq:36}
Z^{1/3} \sqrt{xF(x)} \gg 1\,.
\end{equation}
First, we learn here, that for a given $x$, TF is reliable only if $Z$ is
large enough.
Second, there is information about the regions where TF cannot be trusted for
given~$Z$. 

\begin{figure}[t]
\centering
\includegraphics{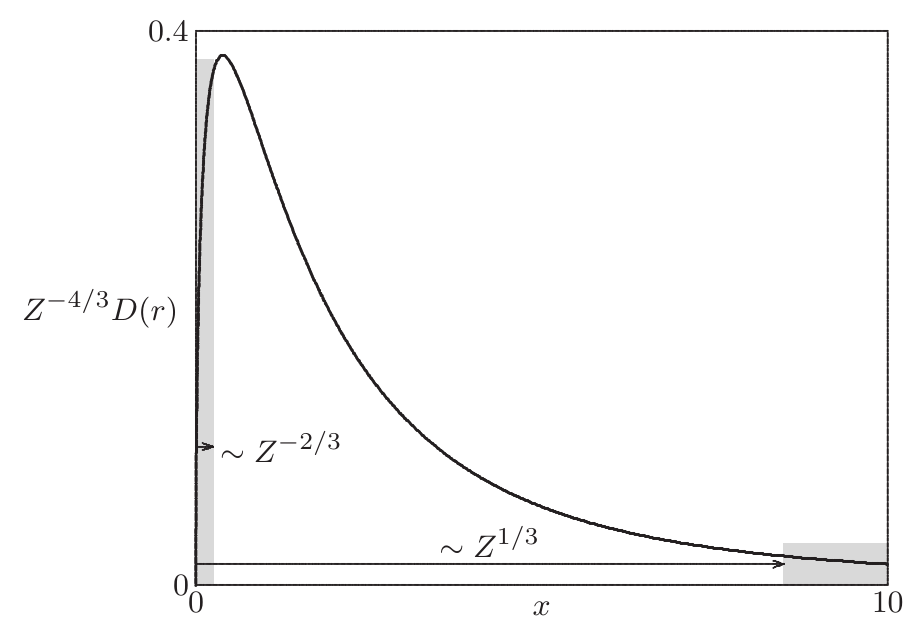}
\caption{Regions of failure of TF, illustrated by the radial density as a
  function of $x$.} 
\label{fig05}
\end{figure}

At short distances, $F(x)$ practically equals unity, and the left-hand side of
\Eq[]{36} is of the order of unity, when $x \sim Z^{-2/3}$, or $r \sim 1/Z$.
Consequently, there is an inner region of strong binding where TF fails.
Then, at large distances, where $F(x ) \sim 1/x^3$, the inequality \Eq[]{36}
is not obeyed, once $x$ is of the order $Z^{1/3}$, or $r\sim 1$.
Now we find the outer region of weak binding to be also described inadequately
by TF.
The entire situation is thus as shown in Fig.~\ref{fig05}.
The two shaded areas are badly treated by TF, whereas the intermediate region
of the atom is dealt with rather accurately.
And, the bigger $Z$, the less important the shaded regions are.
We conclude that (in some sense) TF becomes exact for $Z \rightarrow \infty$.

Nice, but in the real world $Z$ isn't that large, the more so $Z^{1/3}$, which
obviously is the relevant parameter.
It ranges merely from one to roughly five over the whole Periodic Table.
Clearly, modifications aimed at improving TF are called for.

\enlargethispage{-1.0\baselineskip}

\section{Improving TF.  Strongly Bound Electrons}
Despite their relatively small number, the electrons close to the nucleus have
such a large binding energy that the leading correction to the TF energy
formula \Eq[]{32} stems from a better description of these strongly bound
electrons.
This can be achieved rather simply, with a very rewarding outcome.
Here is how it goes.

\begin{figure}[t]
\centering
\includegraphics{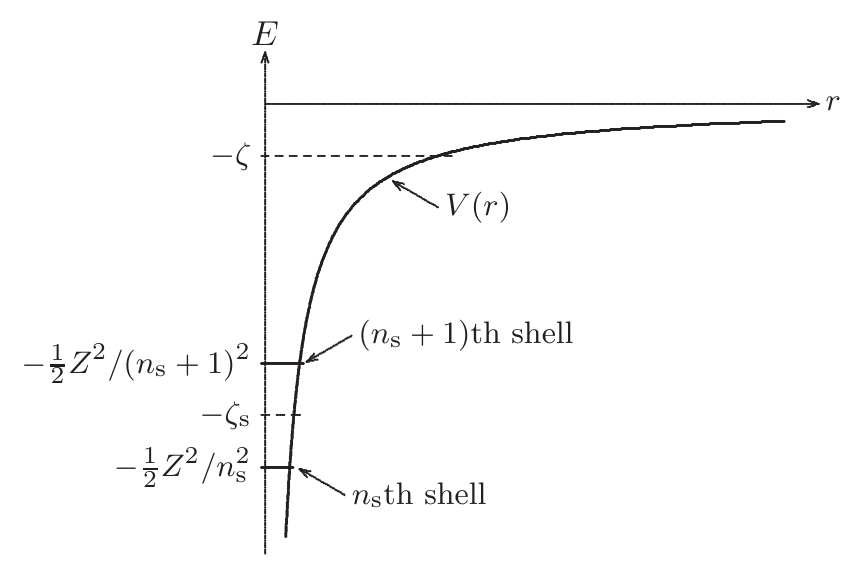}
\caption{Concerning corrections for strongly bound electrons; see text.}
\label{fig06}
\end{figure}

In the region we are talking about now, the vicinity of the nucleus, the
potential is well approximated by \Eq[]{29};
it is the Coulomb potential plus a (small) constant.
In other words, the strongly bound electrons feel practically only the force
of the nucleus, while the interaction with other electrons is negligible.
Consequently, we shall treat the neighborhood of the nucleus as occupied by
NIE, filling a certain number, $\ns$, of Bohr shells.
Formally, the one-particle energy spectrum is divided at a binding energy
$\zetas$ which separates the strongly bound electrons from the rest, see
Fig.~\ref{fig06}.
This $\zetas$ is, of course, not a uniquely defined physical quantity. But it
is not arbitrary.
First, $\zetas$ denotes a binding energy that is large on the TF scale, but
small on the Coulomb scale, because we do not want to correct for just the 1s
shell, but for all relevant Bohr shells.
Symbolically this means
\begin{equation}\label{eq:37}
Z^{4/3} \ll \zetas \ll Z^2\,.
\end{equation}
Second, $\zetas$ is related to the number of shells, $\ns$, that are treated
specially.
So $\zetas$ has to be sandwiched by the binding energies of the $\ns$th and
the $(\ns+1)$th shell.
This can be expressed by
\begin{equation}\label{eq:38}
\ns < \frac{Z}{\sqrt[]{2\zeta_s}} < \ns + 1 \,,
\end{equation}
which exhibits the combination of $Z$ and $\zetas$ that will be relevant in a
moment.
The union of \Eq[]{37} and \Eq[]{38} shows that the total number of
specially treated strongly bound electrons, $\Ns$ [$\sim \ns^3$, cf.\ \Eq{1}],
is a small fraction of all electrons, 
\begin{equation}
\Ns \ll Z\,.
\end{equation}
To obtain the change in the binding energy generated by the improved
description of the innermost electrons, we have to do two things: remove the
wrong TF value, which is $Z^2\bigl(Z/\sqrt{2\zetas}\bigr)$, and then add the
correct Bohr energy of \Eq{2}.
Thus the correction to the binding energy is
\begin{equation}\label{eq:40}
\Delta(-E) = - Z^2 \frac{Z}{\sqrt{2\zetas}} + Z^2\ns \,.
\end{equation}
This looks ambiguous, inasmuch as it contains $\ns$ and $\zetas$, both being
quantities that lack a unique value.
However, for fixed $\ns$, the possible range for $\zetas$ is given in
\Eq[]{38}, so that averaging over this range assigns the value
$-Z^2\bigl(\ns+\half\bigr)$ to the first term on the right of \Eq[]{40}.
This implies
\begin{equation}\label{eq:41}
\Delta(-E) = - \frac{1}{2} Z^2\,.
\end{equation}
What we have found is the next term in the energy formula; it now reads
\begin{equation}\label{eq:42}
\frac{-E}{\frac{1}{2}Z^2} = 1.537 Z^{1/3} - 1\,.
\end{equation}
Since the strongly bound electrons involved in this correction are basically
noninteracting, it could have been anticipated that the additional term is
identical with the respective one in the formula for NIE, \Eq{5}.

\begin{figure}[t]
\centering
\includegraphics{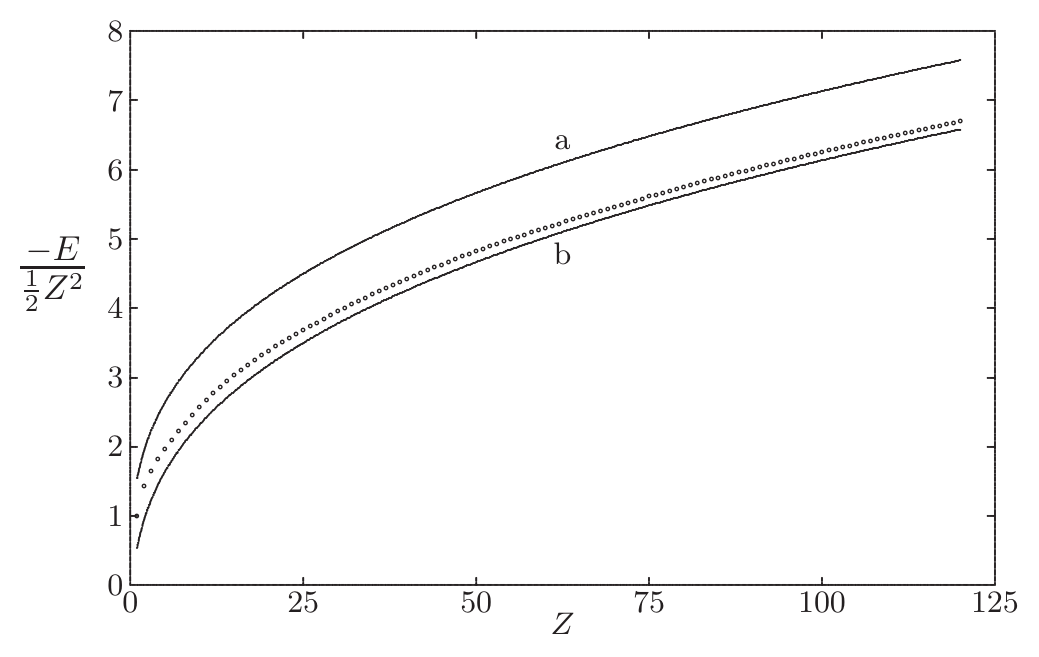}
\caption{Comparison between binding energies as predicted by HF (circles); TF
  (curve a); TF with corrections for strongly bound electrons (curve b).} 
\label{fig07}
\end{figure}

Again we compare with the HF energies.
Figure~\ref{fig07} also gives a plot of the previous TF curve to emphasize the
significant improvement.

\section{Improving TF.  Quantum Corrections and Exchange}
The corrections we discuss next come from the main body of the electrons.
First we remark that \Eq{14} is in error to the extent that the quantum
effects introduced in \Eq[]{33} are significant.
So we now consider corrections associated with the finiteness of $\grad V$.
Inasmuch as this is a vectorial quantity, the leading energy correction is of
second order\cite{fn4} in the parameter of \Eq[]{33},
$\boldsymbol{|}\lambda\grad V\boldsymbol{|} / \boldsymbol{|}V\boldsymbol{|}$,
and therefore produces an energy change $\sim Z^{5/3}$.
For details the reader is referred to Ref.~\citen{fn5}, from which we quote
the leading quantum correction to the energy, 
\begin{equation}\label{eq:43}
  \Delta E_{\mathrm{qu}} = - \frac{1}{18 \pi^3} \int (\D\vec{r})\,
  \bigl[-2(V+\zeta)\bigr]^2\,.
\end{equation}
The derivatives of the potential that occur initially have been removed both
by partial integrations and by utilizing the TF equation for $V$, \Eq{17}.
For neutral atoms, the corresponding supplement to \Eq[]{42} is now obtained
by inserting the TF potential (and $\zeta = 0$, of course) into \Eq[]{43},
which leads to 
\begin{equation}\label{eq:44}
  \Delta E_{\mathrm{qu}} =
  - \frac{Z^{5/3}}{16a^2} \int_0^\infty \D x\, F(x)^2 
  = \frac{2}{11} \bigl(-0.2699Z^{5/3}\bigr)\,.
\end{equation}
[The integral has the numerical	value 0.6154.]
Again, as for the leading $Z^{7/3}$ term, this $Z^{5/3}$ contribution is
roughly two thirds of the $Z^{5/3}$ term for NIE in \Eq{3}.

\begin{figure}[t]
\centering
\includegraphics{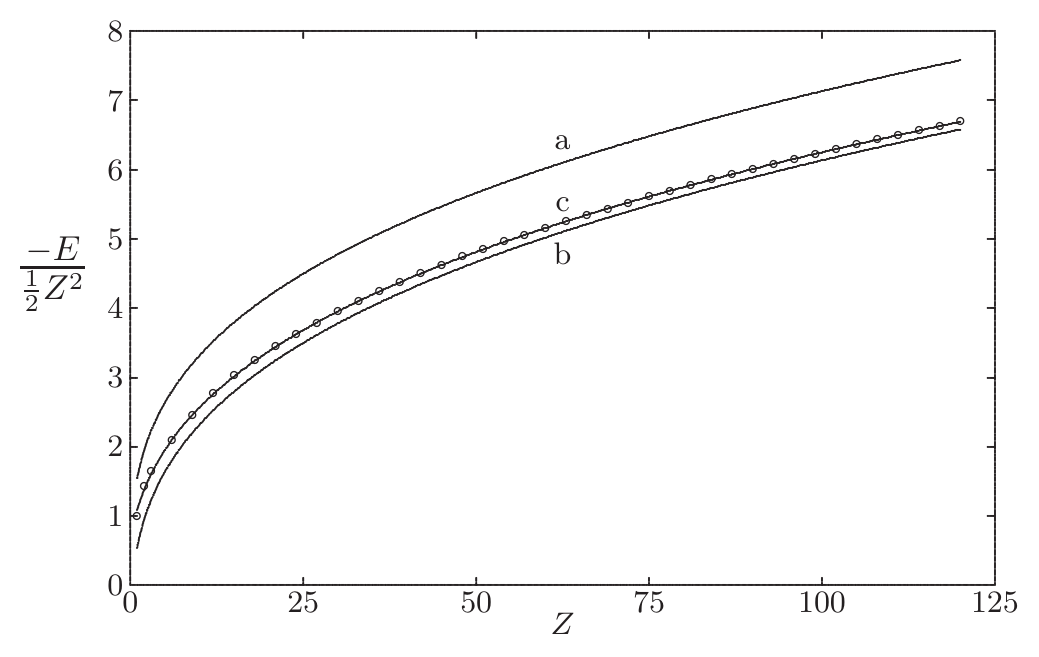}
\caption{Comparison between binding energies as predicted by HF (circles;
  $Z=1,2,3,6,9,\ldots, 120$ shown only); TF (curve a); TF with corrections for
  strongly bound electrons (curve b) ; the statistical binding energy formula
  \Eq[]{46} (curve c).} 
\label{fig08}
\end{figure}

There is a second effect that also produces a $Z^{5/3}$	correction to the
energy --- the exchange interaction of the electrons.
In contrast with the electrostatic interaction energy of each electron with
the other electrons, constituting $Z$ electrons at a distance $\sim Z^{1/3}$,
which is of order $Z/Z^{-1/3}=Z^{4/3}$, exchange is limited to electrons with
overlapping wave functions at a distance
$\sim \lambda \sim 1/\boldsymbol{|}V\boldsymbol{|}^{1/2} \sim Z^{-2/3}$;
thus the exchange energy of each electron is of the  order $1/Z^{-2/3} =
Z^{2/3}$.
The explicit result of the calculation is\cite{fn5,fn6}
\begin{eqnarray}\label{eq:45}
  \Delta E_{\mathrm{ex}} &=& - \frac{1}{4\pi^3} \int (\D\vec{r})\,
                             \bigl[-2(V+\zeta)\bigr]^2 \nonumber
  \\&=& \frac{9}{2} \Delta E_{\mathrm{qu}}
  =  \frac{9}{11}\bigl(-0.2699 Z^{5/3}\bigr)\,.
\end{eqnarray}
It supplements (44) and yields the final statistical energy formula
\begin{equation}\label{eq:46}
\frac{-E_\mathrm{stat}}{\frac{1}{2}Z^2} = 1.537 Z^{1/3} - 1 + 0.5398 Z^{-1/3}.
\end{equation}
A plot of the successive levels of approximation is given in
Fig.~\ref{fig08}.
The marvelous agreement of \Eq[]{46} with HF --- the curve goes right through
the circles --- is a great triumph of the statistical method.
One now understands why these atomic binding energies are so regular.
They are a property of the ensemble of electrons, no individuality is (yet!)
recognizable. 
 
\section{History}
The three terms of \Eq[]{46} are associated with certain names, and we welcome
the opportunity to give a brief historical account.
The subject started with Thomas's paper of November 1926.\cite{Thomas-Fermi}
He could have, but did not derive the leading term of the binding energy
formula.
The first to write down \Eq{32}, in July 1927, was Milne\cite{fn7} who ---
being an astrophysicist --- recognized the similarity of the TF equation
\Eq[]{24} with Emden's equation for spheres of polytropic perfect gases, held
together by gravitation.
Milne's numerical factor was about twenty percent too small, which
accidentally improved the agreement with the then available experimental
data.
Fermi's first paper on the statistical theory of atoms was published in
December 1927.\cite{Thomas-Fermi}
It contains a remarkably good numerical solution for $F(x)$ [he calls it
$\varphi$]; e.g., the initial slope $B$ is given as $1.58$.
Fermi also noticed the connection between the total binding energy and this
constant, so that he can claim fatherhood of \Eq{31}. His numerical factor is,
of course, much better than Milne's --- only half a percent short of the
modern value.
We are told that Fermi was unaware of Thomas's work until late in 1928, ``when
it was pointed out to him by one (now unidentified) of the foreign
theoreticians visiting Rome.''\cite{fn8}
There are two probable candidates for this anonymous person:
Bohr and Kramers, whose encouragement is acknowledged by Thomas in his
paper.\cite{Thomas-Fermi} 

The credit for the first highly accurate calculation of $F(x)$ belongs to
Baker.\cite{fn9}
His work was published in 1930, long before the age of high-speed computers,
and contains a value for $B$ which is exact to 0.03\%.
We honor Baker by assigning his initial to this number.

Now to the next term in \Eq[]{46}, the correction for strongly bound
electrons.
While it has, of course, always been recognized how badly the innermost
electrons are represented by TF, it would take the surprisingly long time of
25 years until Scott came up with the energy correction of \Eq{41}, in
1952.\cite{fn10}
However, his derivation --- he calls it a ``boundary effect'' and treats it
accordingly --- has not been widely accepted.
The general feeling concerning Scott's correction was that ``it seems
difficult to give a completely clearcut demonstration of the
case.''\cite{fn11}
This was delivered --- in the spirit of the treatment reported above --- by
one of us in 1980,\cite{fn12} another 28 years later.
Recently, we showed an elegant derivation of Scott's term by making use of the
scaling properties of TF with corrections for strongly bound
electrons.\cite{fn13} 

Scott, in the very same paper,\cite{fn10} was also the first to give a
$Z^{5/3}$ term in the energy formula.
However, being unaware of the quantum corrections, he considered merely the
exchange contribution of \Eq{45}, thus accounting for nine elevenths of the
last term in \Eq[]{46}.
Again it took many years before, in 1981, the quantum correction,
Eqs.~(\ref{eq:43}, \ref{eq:44}), was evaluated by one of us.\cite{fn5}
From then on, the statistical energy formula was known.\cite{fn14}
Of course, there has been important work on extensions of TF by other
authors.
The exchange interaction was first considered by Dirac, as early as
1930.\cite{fn15}
[He was possibly reacting to a remark by Fermi at the end of a talk presented
at a 1928 conference in Leipzig,\cite{fn16} which Dirac also attended.]
But Dirac did not deal with exchange \emph{energy}, just with the implied
modification of the TF equation.
An expression for this energy, equivalent to \Eq[]{43}, was first given by
Jensen in 1934,\cite{fn17} who also on this occasion corrected for an
inadvertance of Dirac, whose exchange effect was too large by a factor of 2.
However, there is no doubt that it was Scott who for the first time evaluated
the exchange energy perturbatively, arriving at \Eq[]{45}.
Maybe both Dirac and Jensen were just thinking that one should not talk about
the second correction before the first one is known~\ldots 

The first attempt at including the nonlocality of quantum mechanics was
performed by v.~Weizs\"{a}cker in 1935.\cite{fn18}
He derived a correction to the kinetic energy which has the serious drawback
that it cannot be evaluated in perturbation theory --- the outcome would be
infinite.
From our investigations of quantum corrections\cite{fn5,fn6} it has become
clear that a consistent treatment requires a simultaneous, correct handling of
the strongly bound electrons.
Why didn't Scott do exactly that?
There are two reasons.
First, Scott's ``boundary effect'' theory of the vicinity of the nucleus
cannot be directly implemented into the energy functional.
And second, the language used by v.~Weizs\"{a}cker,  Scott, and others is
based on the electron density as the fundamental quantity, whereas these
problems are most conveniently discussed by giving the potential the
fundamental role.

\begin{figure}[t]
\centering
\includegraphics{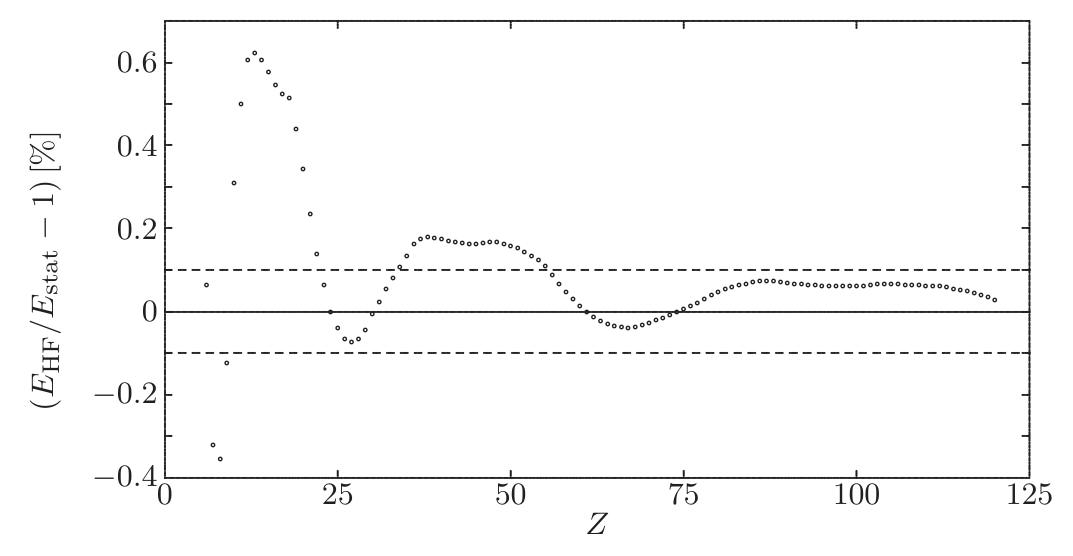}
\caption{Relative deviation, in \%, between HF binding energies and the
  statistical formula \Eq[]{46}.} 
\label{fig09}
\end{figure}

\section{Shell Effects}
\begin{figure}[b]
\centering
\includegraphics{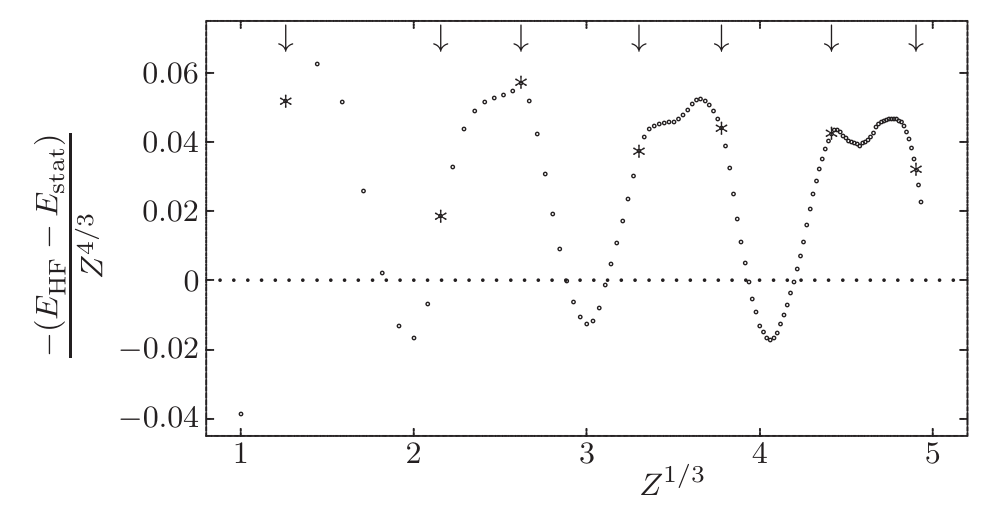}
\caption{Absolute deviation between HF binding energies and the statistical
  formula \Eq[]{46}.
  Stars mark the location of inert gas atoms, and the arrows point to them.} 
\label{fig10}
\end{figure}

Although we were justifiably pleased with the striking agreement of the
statistical curve and the HF crosses in Fig.~\ref{fig08}, the story is not yet
finished. 
Let us now look at this plot as though through a microscope.
Figure~\ref{fig09} shows the relative deviation between \Eq{46} and HF.
One sees that from $Z = 22$ on, the 0.2\% level is reached;
after $Z = 56$ the accuracy is better than 0.1\%.
There is an obvious, steady increase in agreement as $Z$ becomes larger, but
this is accompanied by interesting fluctuations.
It turns out that Fig.~\ref{fig09} is not the appropriate way of looking at
these oscillations.
Instead of the relative difference between HF and \Eq{46} we now present the
absolute deviation, divided by $Z^{4/3}$, which is the anticipated order of
the next term in the binding energy formula, and plotted not as a function of
$Z$ itself, but against $Z^{1/3}$ which is the significant measure of the
number of electrons.
This is Fig.~\ref{fig10}.
We are confronted with an unexpectedly regular oscillatory curve that is not
only well defined for large $Z$ but even reaches all the way down to hydrogen,
$Z = 1$.
Can the statistical approach be employed at all to explain such a
behavior?
Well, although oscillatory, the $Z$-dependence is still smooth and certainly
not irregular.
However, before embarking on a calculation, we first have to gain a
qualitative physical understanding of the underlying phenomena. 

Everybody's immediate reaction to Fig.~\ref{fig10} is that the oscillations
are the filling of atomic shells.
If this were true, surely the atoms with closed shells, the inert gases ---
He~($Z = 2$), Ne~(10), Ar~(18) , Kr~(36) , Xe~(54), Rn~(86), and another one
with ${Z = 118}$ for which the chemists have not yet invented a name --- would
sit on prominent places of the curve?
Figure~\ref{fig10} shows these locations, and on first sight the inert gas
atoms do not seem to follow any pattern related to the oscillatory curve.
They are, however, also not randomly distributed over the oscillatory curve,
but show a clear tendency to be close to its maxima.
We infer, therefore, that there is a connection between the energy oscillation
and the existence of closed atomic shells. 
These \emph{two} phenomena are manifestations of \emph{one} underlying
physical effect.  

\begin{figure}[t]
\centering
\includegraphics{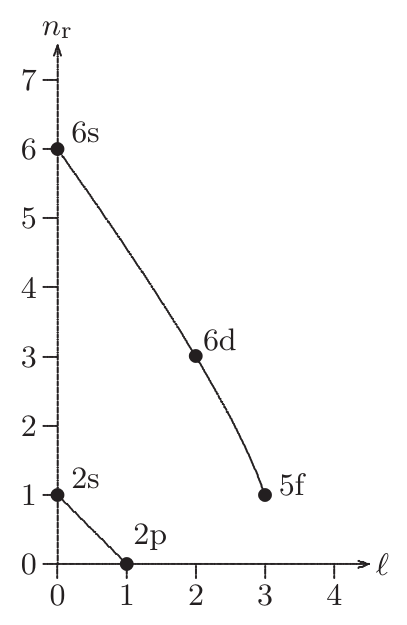}
\caption{Energetic degeneracy in a large atom.}
\label{fig11}
\end{figure}

To answer the question what effect that is, let us recall the origin of atomic
shells.
The reason for their being is the existence of quantum numbers in a
spherically symmetric potential: angular quantum number $\ell$, and radial
quantum number $\nr$.
But $\ell$ and $\nr$ alone would not account for shells; we also need the fact
of energetic degeneracy.
States with differing quantum numbers may have almost the same binding
energy.
This is, of course, familiar for the Coulomb potential where the energies
depend only on the principal quantum number ${n = \nr + \ell + 1}$, leading to
the $2n^2$ fold degeneracy that we have made use of in \Eq[Eqs.~]{1} and
\Eq[]{2}.
Thus, in atoms containing NIE, the maximal radial quantum number and the
maximal angular quantum number are equal.
Not so in the real world, where the ratio of the two is roughly $2:1$
[uranium, e.g., possesses 7s electrons ($\nr=6$) and 5f electrons ($\ell=3$)].
The degeneracy of the weakly bound outermost electrons is certainly not of
Coulombic type.
We can learn more about it from another look at the Periodic Table, this time
at the last row.
There the 7s, 6d, and 5f electrons are filled in, but not in a given order,
instead they compete with each other --- a sure sign of degeneracy.
In an $\ell$-$\nr$ diagram, Fig.~\ref{fig11}, these three states do not lie on
a straight line; degenerate states are connected by bent curves which are the
steeper, the larger $\ell$ is.
Deep inside the atom, we expect Coulombic degeneracy for the strongly bound
electrons.
In this situation, states with the same binding energy do lie on a straight
line in the $\ell$-$\nr$ diagram.
In Fig.~\ref{fig11} this is illustrated by the 2s and the 2p state.

It is clear that a theoretical description of the oscillations in
Fig.~\ref{fig10} must be based on a detailed energetic treatment of those few
electrons with least binding energy.
This view is supported by the relative size of the effect we are looking for,
which is of order $1/Z$ as compared to the leading $Z^{7/3}$ term, or, like
one electron compared to the totality of $Z$ electrons. 

After these preparatory remarks, it is no surprise that we now attempt to
evaluate the effective (i.e., including $\zeta$) single particle energy $E_1$
of \Eq{9} by performing the sum over the quantum numbers,
\begin{equation}\label{eq:47}
E_1=\sum_{\nr,\ell}2(2\ell +1)(E_{\nr ,\ell}+\zeta)\,\eta(-E_{\nr,\ell}-\zeta)\,.
\end{equation}
Here we exhibit the spin and angular momentum multiplicity, and have the step
function select those states with ${E_{\nr,\ell} < -\zeta}$.
We shall relate the individual energies to the potential $V$ and the quantum
numbers, not through the eigenvalue problem associated with Schr\"{o}dinger's
equation, but by means of the semiclassical quantization rule, 
\begin{equation}\label{eq:48}
  \nr + \frac{1}{2} = \frac{1}{\pi} \int \D r\,
  \Bigg[2\biggl( E_{\nr,\ell}-V-\frac{\bigl(\ell+\frac{1}{2}\bigr)^2}{2r^2}
  \biggr)\Bigg]^{1/2}\,,
\end{equation}
which is usually derived by the WKB method.
Equation~(\ref{eq:48}) is known to produce the exact energy eigenvalues for a
few simple potentials, notably the Coulomb and the oscillator potential.
The exactness for Coulombic potentials is significant, since it assures
correct treatment of the strongly bound electrons.
For other ``smooth'' potentials, \Eq[]{48} gives very good approximate values
for $E_{\nr,\ell}$.
Certainly, for our purposes, \Eq{48} is good enough.
But even with the simplifications provided by employing \Eq[]{48}, the double
sum of \Eq[]{47} is not easily evaluated.
The main reason for that is the implicit definition of $E_{\nr,\ell}$ in
\Eq[]{48} where it is the radial quantum number that is expressed as a
function of $\ell$ and $E_{\nr,\ell}$. 

Both quantum numbers are accompanied by an added $\frac{1}{2}$ in \Eq{48}; we
shall therefore simplify matters by introducing new variables according to 
\begin{equation}\label{eq:49}
  \nu \equiv \nr + \frac{1}{2}\,,\quad
  \lambda \equiv \ell + \frac{1}{2}\,,\quad
  \mathcal{E}_{\nu, \lambda} \equiv E_{\nr,\ell}\,,
\end{equation} 
so that now
\begin{equation}\label{eq:50}
  \nu = \frac{1}{\pi} \int \frac{\D r}{r}
  \bigl[ 2r^2 (\mathcal{E}_{\nu,\lambda} - V) - \lambda^2 \bigr]^{1/2}\,.
\end{equation}
In both equations, \Eq[]{48} and \Eq[]{50}, the domain of integration is the
classically allowed region where the argument of the square root is positive.

\begin{figure}[t]
\centering
\includegraphics{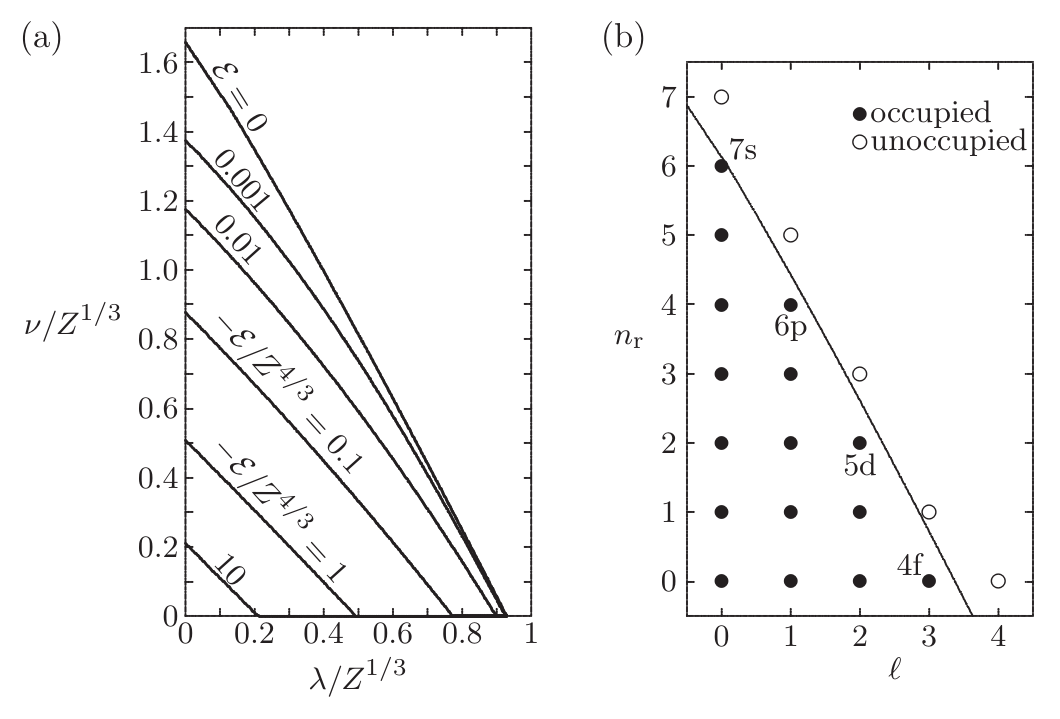}
\caption{(a) Energetic degeneracy in the TF potential. %
         (b) TF prediction for occupied states in Ra ($Z = 88$).}
\label{fig12-13}
\end{figure}

Before proceeding with our investigations of Eq. (47) let us make sure
that the TF potential can be expected to be useful here. Its insertion into
(50) produces the following ratio of maximal values for $\nu$ and $\lambda$:
\begin{equation}\label{eq:51}
  \frac{  \nu (\mathcal{E}=0, \lambda=0) }{ \lambda (\mathcal{E}=0, \nu=0)  }
  = \frac{ 1.659 \, Z^{1/3} }{ 0.928\, Z^{1/3}} = 1.79\,,
\end{equation}
which is in reasonable agreement with the corresponding number for a large
atom, e.g., uranium,
\begin{equation}
  \frac{ \nu_\mathrm{max} }{ \lambda_\mathrm{max} }
  = \frac{ 6+\frac{1}{2}}{3+\frac{1}{2}} = 1.86\,.
\end{equation}
In Fig.~\ref{fig12-13}(a) we see $\nu/Z^{1/3}$ as a function of
$\lambda/Z^{1/3}$ for several $\mathcal{E}$, demonstrating the different
character of degeneracy for small and large binding energies;
a plot that has a striking similarity with Fig.~\ref{fig11}.
This becomes even more
convincing when we ask for the specific states available in a large atom, say
radium, $Z = 88$.
Figure~\ref{fig12-13}(b) shows perfect agreement between the TF prediction and
experimental observations;
the $\mathcal{E} = 0$ curve of Fig.~\ref{fig12-13}(a) separates the occupied
states from the unoccupied ones, selecting exactly those that are
spectroscopically known to be available. 
This is another way of presenting the results of Fermi's application of the
statistical theory of atoms to the systematics of the Periodic Table, the
great historical triumph of TF, published in his second paper on the subject
in January 1928.\cite{fn19} 

\begin{figure}[t]
\centering
\includegraphics{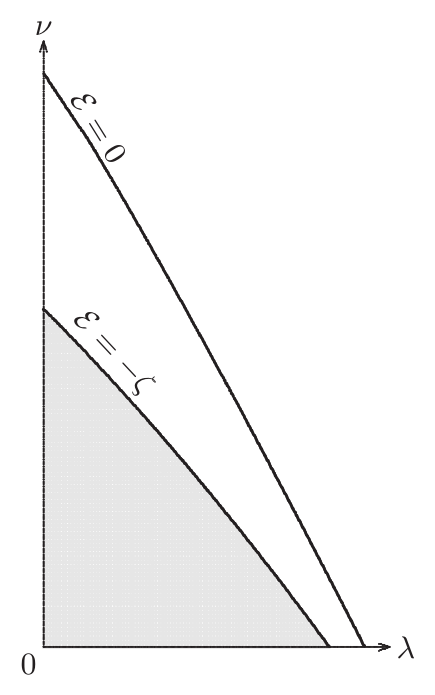}
\caption{Domain of integration in \Eq{54} is shaded area.}
\label{fig14}
\end{figure}

With this reassurance, the discussion of \Eq{47} is now continued.
A more useful way of writing this sum over quantum numbers employs $\nu$ and
$\lambda$ of \Eq[]{49} and replaces the summation, e.g., over $\ell$, by an
equivalent integration over $\lambda$, with the aid of the Poisson sum
formula, 
\begin{eqnarray}\label{eq:53}
  \sum^\infty_{\ell=0} 2(2\ell+1) A_{\ell + \frac{1}{2}}
  &=& 4 \int^\infty_0 \D\lambda\, \lambda \sum^\infty_{\ell=-\infty}
      \delta\bigl( \lambda - \ell - \thalf \bigr) A_\lambda \nonumber \\
  &=& 4 \sum^\infty_{k=-\infty} (-1)^k
      \int^\infty_0 \D\lambda\,\lambda \,\Exp{\I2\pi k \lambda} A_\lambda\,.
\end{eqnarray}
This, with the analogous procedure for $\nr$ and $\nu$, turns \Eq[]{47}	into
\begin{equation}\label{eq:54}
  E_1 = 4 \sum^\infty_{k,j=-\infty} (-1)^{k+j} \int^{\lambda_\mathcal{E}}_0
  \D\lambda \, \lambda \, \Exp{\I2\pi k\lambda}
  \int^{\nu (\mathcal{E}=-\zeta,\lambda)}_0 \D\nu \; \Exp{\I2\pi j \nu}
  (\mathcal{E}_{\nu,\lambda} + \zeta)\,.
\end{equation}
We got rid of the step function by making the limits of integration explicit;
the domain of integration is the area between the $\nu, \lambda$ axes and the
curve belonging to $\mathcal{E} = - \zeta $ in Fig.~\ref{fig14}.
So far we have done nothing but rewrite \Eq{47}.
Now we shall illustrate this structure by picking out the $j=0$ terms.
In other words, we concentrate on $\lambda$ oscillations only, disregarding
$\nu$ oscillations, which corresponds to a simplified picture in which only
angular momentum is quantized, not the radial motion. 
We call this $\ell$TF, short for $\ell$-quantized Thomas--Fermi.

The absence of the exponential in the $\nu$ integral for $j = 0$ enables us to
change variables from $\nu$ to $\mathcal{E}$ (for fixed $\lambda$) which is
desirable in view of the implicit definition of $\mathcal{E}_{\nu,\lambda}$ in
\Eq{50}.
Here is how it goes:
\begin{equation}
  \D\nu\, (\mathcal{E} + \zeta)
  = \D[\nu\,(\mathcal{E}+\zeta)]-\nu\,\D\mathcal{E}\,;
 \end{equation}
the total differential has zero value at both limits of the integral, and $\nu$
possesses a simple algebraic dependence on $\mathcal{E}$, allowing further
integration, 
\begin{equation}
  \D\nu\,(\mathcal{E}+\zeta) \rightarrow \D\mathcal{E}
  \frac{\partial}{\partial\mathcal{E}}
  \Biggl( - \frac{1}{3\pi} \int \frac{\D r}{r^3}
    \bigl[ 2r^2 (\mathcal{E}-V)-\lambda^2 \bigr]^{3/2} \Biggr)\,.
\end{equation}
The $\ell$TF result for $E_1$ is then
\begin{equation}
  E_1^{\ell TF} = - \frac{4}{3 \pi} \sum_{k=-\infty}^{\infty} (-1)^k
\int \D\lambda \, \lambda \, \Exp{\I2\pi k \lambda}
\int \frac{\D r}{r^3} \bigl[ 2r^2 (-\zeta - V)-\lambda^2 \bigr]^{3/2}\,.
\end{equation}
This equation was, in some sense, known to Hellmann in 1936,\cite{fn20} but
certainly not in this form.
His formula used the original sum over $\ell$ [easily obtained by reversing
\Eq[]{53}], and was expressed in terms of densities, one for each value of
$\ell$. 
This was a clumsy way of writing it, which unfortunately kept both Hellmann
and Gomb\'{a}s, who devoted a chapter in his classical textbook\cite{fn21} to
the matter, from realizing the important fact that the $k = 0$ contribution to 
this sum is just the TF expression \Eq[]{16}, once the $\lambda$ integration
is carried out.\cite{fn22}

\begin{figure}[t]
\centering
\includegraphics{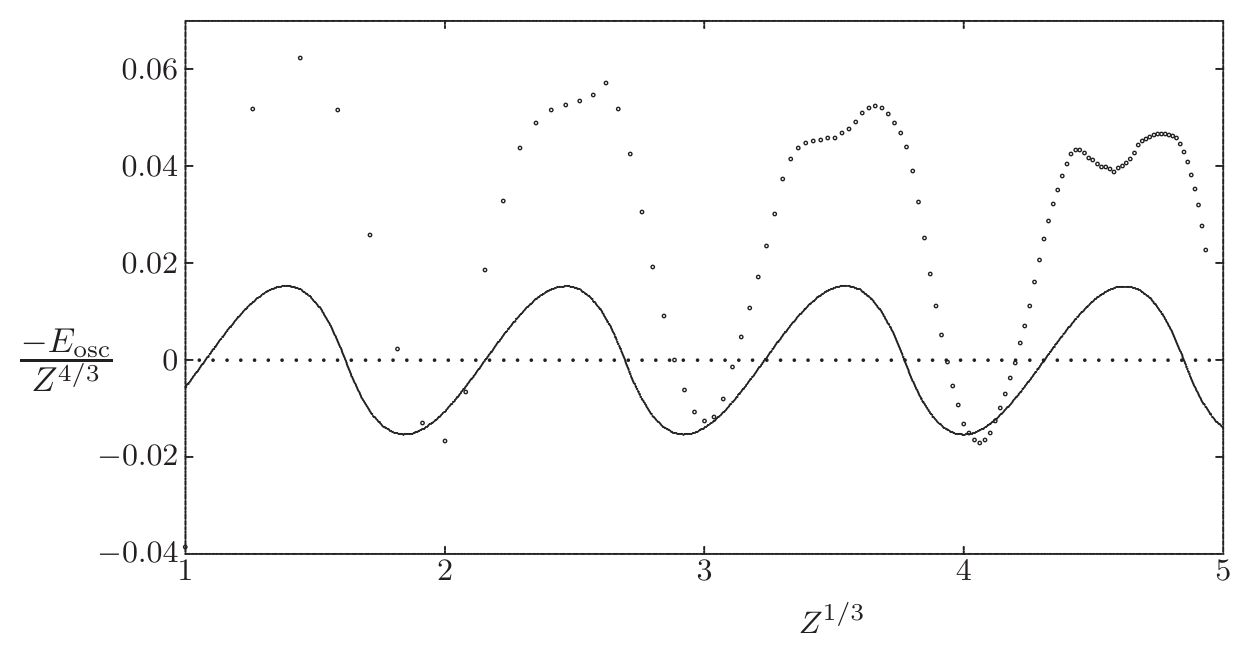}
\caption{Comparison of HF oscillations (circles) %
         with leading $\ell$TF oscillations (solid line).} 
\label{fig15}
\end{figure}

With that essential piece of information, we split $E_1^{\ell \mathrm{TF}}$
into $E_1^{\mathrm{TF}}$ and the rest, 
\begin{equation}
  E_1^{\ell \mathrm{TF}} - E_1^{\mathrm{TF}}
  = \frac{8}{3\pi} \sum^\infty_{k=1} (-1)^{k-1}
  \int \frac{\D r}{r^3} \int\D\lambda\,\lambda\,
  \bigl[ 2r^2(-\zeta -V)-\lambda^2 \bigr]^{3/2}
  \cos (2\pi k \lambda)\,,
\end{equation}
and, since this right-hand side must be a small correction to TF, we are
justified in using the TF potential for its evaluation.
The leading $\ell$TF oscillation can be easily identified.
To that end, we first replace $\lambda$ by its maximum value times the cosine
of an angle $\theta$, 
\begin{equation}
\lambda = \bigl[ 2r^2 (-\zeta -V)\bigr]^{1/2} \cos \theta\,;
\end{equation}
this leads to
\begin{eqnarray}\label{eq:60}
  E_1^{\ell \mathrm{TF}} - E_1^{\mathrm{TF}}
  &=& \frac{8}{3\pi} \sum^\infty_{k=1} (-1)^{k-1}
      \int^\infty_0 \frac{\D r}{r^3}
       \bigl[ 2r^2(-\zeta -V) \bigr]^{5/2}  \nonumber \\
  && \hphantom{\frac{8}{3\pi} \sum^\infty_{k=1}}
     \times \int^{\pi/2}_0 \D\theta \, \cos \theta \, \sin^4 \theta \,
     \cos (z\cos \theta)
\end{eqnarray}
with
\begin{equation}
z = 2\pi k \bigl[ 2r^2 (-\zeta-V)\bigr]^{1/2}\,.
\end{equation}
Then the asymptotic form of the $\theta$-integral in \Eq{60} for large $z$
[compare \Eq{35}] is employed in identifying the leading oscillatory term, 
\begin{equation}
  E_{\mathrm{osc}}^{\ell \mathrm{TF}}
  = - \frac{1}{\pi^3} \sum^\infty_{k=1} \frac{(-1)^{k-1}}{k^{5/2}}
  \int^\infty_0 \frac{\D r}{r^3}\bigl[ 2r^2 (-\zeta-V)\bigr]^{5/4}
  \, \cos\bigl( z - \tfrac{1}{4}\pi\bigr)\,.
\end{equation}
The last step is a stationary phase evaluation of the remaining $r$-integral
for $V = V_{\mathrm{TF}}$ and $\zeta=0$.
The explicit form of the resulting leading $\ell$TF oscillation is 
\begin{equation}\label{eq:63}
  E_{\mathrm{osc}}^{\ell \mathrm{TF}}
  = - 0.4805 Z^{4/3}  \sum^\infty_{k=1} \frac{(-1)^k}{(\pi k)^3}
  \sin (2\pi k \lambda_0)
\end{equation}
wherein $\lambda_0$ is the maximal value of  of $\lambda$ [cf.~\Eq[]{51}],
\begin{equation}
\lambda_0 = \max_r \bigl[ 2r^2 (-V_{\mathrm{TF}} ) \bigr]^{1/2} = 0.928 Z^{1/3}\,.
\end{equation}
The high power of $k$ in the denominator in the sum of \Eq[]{63} assures us
that a smooth function is represented by this Fourier series; indeed, it is a
repeated piece of a cubic polynomial.
Let us compare $E_{\mathrm{osc}}^{\ell \mathrm{TF}}$ with the HF oscillations of
Fig.~\ref{fig10}, which is plotted in Fig.~\ref{fig15}.
It looks very encouraging because a number of details are quite right:
first, the overall amplitude factor $Z^{4/3}$;
second, the periodicity $Z^{1/3} \rightarrow Z^{1/3} + 1.08$; third, the phase.

\enlargethispage{1.0\baselineskip}

\begin{figure}[t]
\centering
\includegraphics{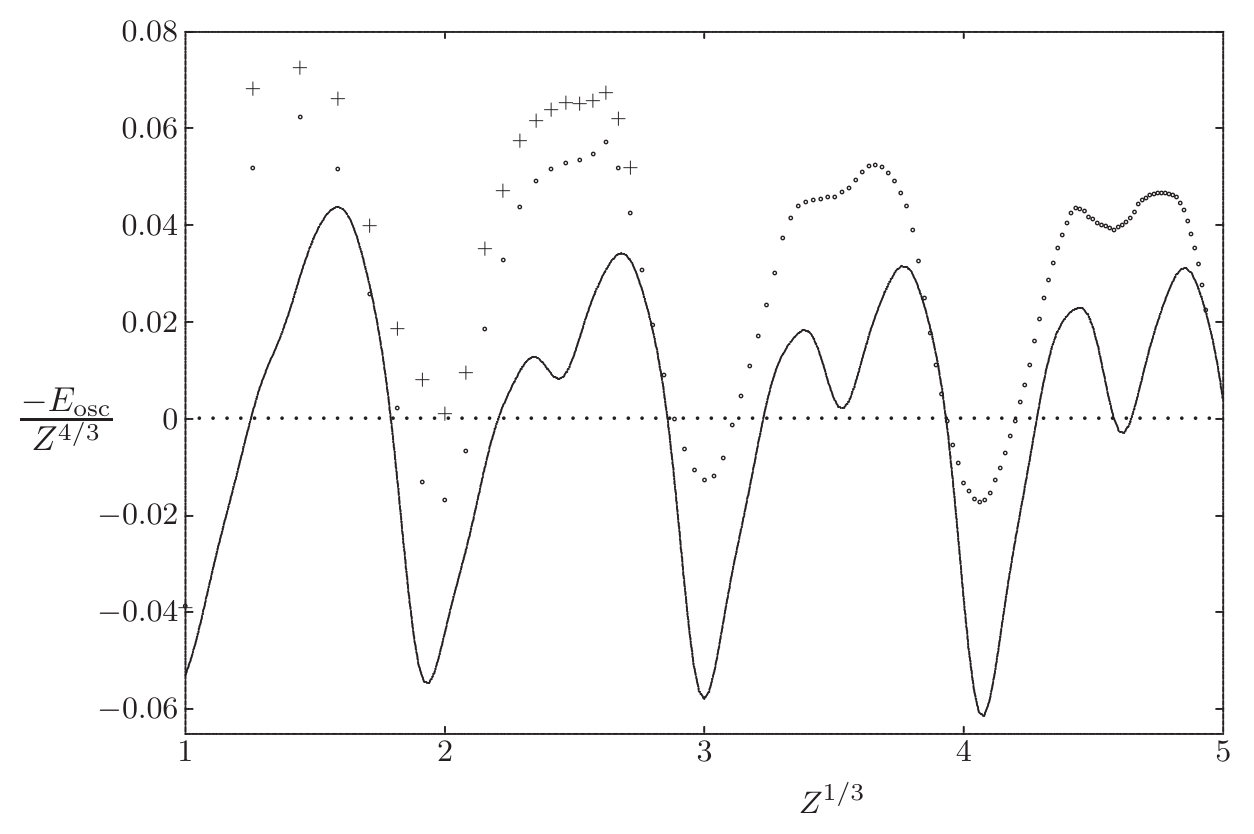}
\caption{Comparison of HF oscillations (circles) with the semiclassical ones
  (solid line) and with experimental data (crosses, corrected for relativistic
  effects).} 
\label{fig16}
\end{figure}

Obviously,  $E_{\mathrm{osc}}^{\ell \mathrm{TF}}$ of \Eq[]{63} is not the
entire oscillation.
It accounts for roughly half the amplitude, but does not show any sign of the
intriguing structure that evolves at the maxima with increasing $Z^{1/3}$.
On the other hand, \Eq{63} was obtained by picking out only the leading
contribution to the $j=0$ term of the double sum in \Eq{54}.
Naturally, a better result should be obtained by evaluating the whole sum.
This is somewhat involved, however, and we have described details
elsewhere.\cite{fn23}
We restrict the present discussion to general aspects.
The analysis shows that one has to pay attention to two major things.
One is that the leading contribution of asymptotic amplitude $\sim Z^{4/3}$
does not suffice, the next-to-leading oscillation ($\sim Z^{3/3}$ for large
$Z$) is also needed; this is reminiscent of the smooth part of the energy
formula, where the leading TF term also did not give satisfactory results.
The other one is the fact that extrapolation from the large-$Z$ domain to the
small-$Z^{1/3}$ region in question has to be done with extreme care; this is
contrary to the situation of the smooth curve, where extrapolating was easy.
Here then is the final semiclassical oscillation: Fig.~\ref{fig16}.

No doubt, our calculation provides a clear understanding of these binding
energy oscillations.
Both the HF curve and the semiclassical one have the same periodicity, the
same amplitude, the same phase, and the same general shape of plain minima,
and maxima with an evolving structure.
True, the HF oscillations show less of this structure, but we never expected
perfect agreement.
[One should not be misled by the obviously missing constant $\sim Z^{4/3}$
that shifts the semiclassical curve down. The calculation concentrated on
oscillatory terms, systematically discarding all smooth contributions.]
While this comparison of two independent theoretical predictions is
entertaining, more to the point is to see what experiment tells.
Total binding energies of atoms have been measured (by stepwise ionization)
for the first 20 members of the Periodic Table only.\cite{fn24}
After correcting for relativistic effects and subtracting the smooth
background of \Eq{46} these experimental values are given by the crosses in
\mbox{Fig.~\ref{fig16}}.
They demonstrate that the oscillations we have been considering are really
present in nature, both period and amplitude being of the theoretically
predicted size.
Neither HF nor the semiclassical curve represent an exact quantitative
description of experiment, while the qualitative agreement is about the same. 

\enlargethispage{1.0\baselineskip}

\section{Concluding Remarks}
We have been focusing on nonrelativistic binding energies throughout this
review.
Naturally, there has been work on relativistic corrections to the statistical
binding energy formula \Eq[]{46}, although the results are not yet quite
satisfactory.\cite{fn12,fn26}
This field remains to be tilled.

Other aspects of the theory concern various properties of atoms, such as
densities, diamagnetic susceptibilities, electric polarizabilities, etc.
Here one encounters different problems.
For the described energy considerations it was sufficient to treat all
modifications of TF in first order perturbation theory (even the
oscillations).
For the investigation of a given atomic system with specified $Z$ and $N$,
this approach is not practicable.
Instead, the corrections for strongly bound electrons, the quantum
improvements, and exchange have to be incorporated consistently into the
energy functional, thus leading to a new differential equation for the
potential [with only a remote resemblance to the TF equation \Eq[]{17}].
This is an entirely different story; we told it recently.\cite{fn6,fn13,fn25}

Then there are generalizations of the statistical theory to other objects than
atoms: molecules, solids, nuclei, even neutron stars, etc.
We do no more than mention them.

\section*{Acknowledgment}
One of us (B.-G.~E.) gratefully acknowledges the generous support by the
Alexander von Humboldt-Stiftung, which granted a Feodor Lynen fellowship.


\begin{thebibliography}{00}
\bibitem{fn3}
The accompanying infinitesimal increase of the number of electrons does not
change the energy because of the null value of $\zeta$, \Eq{19}, as follows
from this implication of \Eq{10}: $\partial E/\partial N=-\zeta$.
\bibitem{fn4}
For consistency, it is also necessary to include a contribution from the
second derivative of the potential; it gives a similar result.
\end{thebibliography}
\end{document}